\documentclass[aps,prb,twocolumn,showpacs]{revtex4}
\usepackage{natbib}
\usepackage{graphics}

\newcommand{\insertgraphics}[2]{
	\begin{figure}
	\begin{center}
	\scalebox{0.8}{\resizebox{\columnwidth}{!}{\includegraphics{#1}}}
	\caption{#2}
	\end{center}
	\end{figure}
}

\newcommand{\insertwidegraphics}[2]{
	\begin{figure}
	\begin{center}
	\resizebox{\columnwidth}{!}{\includegraphics{#1}}
	\caption{#2}
	\end{center}
	\end{figure}
}

\begin{document}

\title{Electron Spectral Functions of Reconstructed Quantum Hall Edges}

\author{A. Melikidze and Kun Yang}
\affiliation{National High Magnetic Field Laboratory,
		1800 E. Paul Dirac Dr.,
		Tallahassee, FL 32310}

\date{November 7, 2003}

\begin{abstract}

During the reconstruction of the edge of a quantum Hall liquid,
Coulomb interaction energy is lowered through the change in
the structure of the edge. We use theory developed earlier by one of
the authors [K. Yang, Phys. Rev. Lett. {\bf 91}, 036802 (2003)]
to calculate the electron spectral functions of a reconstructed edge,
and study the consequences of the edge
reconstruction for the momentum-resolved tunneling into the edge. It
is found that additional excitation modes that appear after the
reconstruction produce distinct features in the energy and momentum
dependence of the spectral function, which can be used to detect the
presence of edge reconstruction.

\end{abstract}

\pacs{73.43.-f, 71.10.Pm}

\maketitle


The paradigm of the Quantum Hall effect (QHE) edge physics is based on an
argument due to Wen,~\cite{Wen} according to which the low-energy edge
excitations are described by a chiral Luttinger liquid (CLL) theory.
One attractive feature of this theory is that due to the chirality,
the interaction parameter of the CLL is often tied to the robust topological
properties
of the bulk and is independent of the details of electron interaction and
edge confining potential; studying the physics at the edge thus offers
an important probe of the bulk physics. It turns out, however, that
the CLL ground state may not always be stable.~\cite{MacDonald-Yang-Johnson,
Chamon-Wen} On the microscopic level,
the instability is driven by Coulomb interactions and leads to
the change of the structure of the edge. This effect has been termed
``edge reconstruction''. One of its manifestations is the appearance of
new low-energy excitations of the edge not present in the original CLL
theory.

Based on the insight from numerical studies of the edge
reconstruction,~\cite{Wan-Yang-Rezayi,Wan-Rezayi-Yang} a field-theoretic
description of the effect has been proposed by one of us.~\cite{Yang}
Remarkably, it provides an
explicit expression for the electron operator in terms of the fields
that describe the low-energy edge excitations after the reconstruction.
This allows one to calculate many observable quantities. In this paper,
we describe the calculation of the spectral function of the electron
in the reconstructed edge. This
function can be probed in tunneling experiments where the electron's
momentum parallel to the edge is conserved
(the so-called momentum-resolved tunneling). Experiments of this kind are
currently being performed.~\cite{Kang,Grayson}
It has also been proposed that momentum-resolved tunneling may be used to
detect the multiple branches of edge excitations of hierarchy
states.~\cite{Zulicke-Shimshoni-Governale}

Our results show that the appearance of the new edge excitations after
the reconstruction modifies the electron spectral function qualitatively.
It leads to the redistribution of the spectral weight away from the
peak corresponding to the original edge mode, and produces singularities
corresponding to the new edge modes.
For simplicity we have focused on the principal Laughlin sequence, 
although generalization to the hierarchy states should be straightforward.
We also propose a particular experimental setup that involves
momentum-resolved electron tunneling which is ideal
for detecting possible edge reconstruction.


\insertgraphics{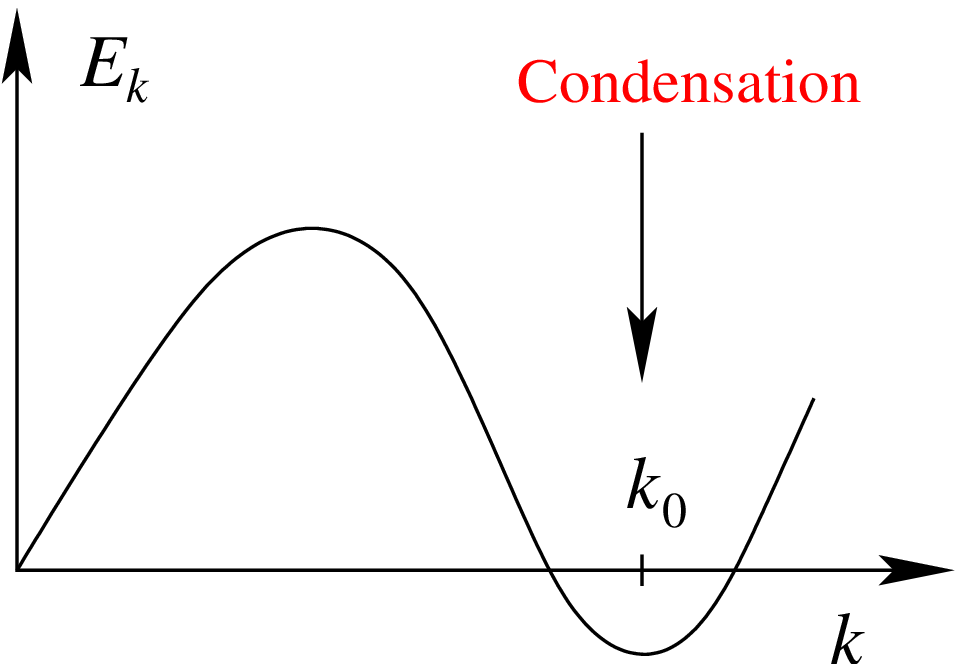}{The spectrum of the edge excitations
after the reconstruction has a minimum below 0. Excitations with momenta
near $k_0$ are condensed.}

Previous numerical studies~\cite{Wan-Rezayi-Yang}
have suggested that the phenomenon of edge
reconstruction can be understood as an instability of the original
edge mode described by the CLL theory.
This instability occurs as a result of increasing curvature of the  
edge spectrum as the edge confining potential softens. The spectrum curves
down at high values of momenta until it touches zero at the transition point
(see Fig.~1). This signals an instability of the ground state as
the edge excitations begin to condense at finite momentum $k_0$.
Such condensation implies the appearance
of a bump in the electron density localized in real space at
$y_0 = l^2 k_0$, where $y_0$ is the coordinate normal to the edge and
$l$ is the magnetic length. These condensed excitations form
a superfluid (with power-law correlation) which possesses a
neutral ``sound'' mode that can propagate in both directions.

The CLL scenario outlined above is relevant to the case of sharp
edge confining potential. In the opposite limit of soft confining
potential, the description of the edge in terms of intermittent
compressible and incompressible stripes has been developed by
Chklovskii et al.~\cite{Chklovskii} In the same limit, emergence of new
edge excitation modes was predicted.~\cite{Aleiner,Balev}
We believe that the description of the
crossover between the two limits should be possible by extending the
present scenario to multiple edge reconstruction transitions, each
characterized by a different momentum $k_i$, at which the instabilities
occur. We comment on this possibility throughout the text.

Following Ref.~\onlinecite{Yang}, we introduce two slowly varying fields
$\phi_1$ and $\varphi$ which describe the original charged edge mode
and the pair of the two new neutral modes, respectively. The total
action of the reconstructed edge of the FQHE at filling fraction
$\nu=1/m$ is: $S=S_1+S_\varphi+S_{\rm int}$, where
\begin{eqnarray}
\label{Action1}
S_1 &=& \frac{m}{4\pi}\int dt\,dx\,\left[\partial_t\phi_1\partial_x\phi_1
	-v(\partial_x\phi_1)^2\right],\\
\label{Action2}
S_\varphi &=& \frac{K_1}{2}\int dt\,dx\,
	\left[\frac{1}{v_\varphi}(\partial_t\varphi)^2
		-v_\varphi(\partial_x\varphi)^2\right],\\
\label{Action3}
S_{\rm int} &=& -K_2\int dt\,dx\, (\partial_t\varphi)(\partial_x\phi_1).
\end{eqnarray}
Here, $S_1$ is the action of the original chiral charged mode with velocity
$v$ , $S_\varphi$ is
the action of the new neutral modes with velocities $v_\varphi$,
$S_{\rm int}$ describes the interaction
between them. The theory, developed in Ref.~\onlinecite{Yang}, predicts:
$K_1 \sim K_2\sim 1$, $v\gg v_\varphi$.
The latter inequality comes from the fact that the Coulomb interaction
boosts the velocity of the charged mode, but not those of neutral ones.

The action in Eqs.~(\ref{Action1},\ref{Action2},\ref{Action3})
describes the simplest situation
where edge excitation condensed in the vicinity of a single point $k_0$.
Such condensation may also occur at multiple points,
producing a set of pairs of additional neutral modes. The action
in that case is a straightforward extension of
Eqs.~(\ref{Action2},\ref{Action3}). We shall comment on the effect
of such multiple edge reconstruction below.

The theory~\cite{Yang} also
provides an explicit expression for the electron operator:
\begin{eqnarray}
\Psi = \exp\left\{im\phi_1 + \rho\cos(k_0x+\varphi)\right\}.
\end{eqnarray}
It has two parts: The first part contains $\phi_1$ and describes an
electron as being made up of the excitations of the original chiral edge
mode. The second part contains $\varphi$ and corresponds to the excitation
of the neutral modes. The constant $\rho$ is proportional to the density of
the new condensate and thus rises from zero at the reconstruction
transition.

Our goal is to calculate the electron spectral function. We begin by
evaluating the Green's function in real space and imaginary-time
representation: $g(x,\tau)=-i\langle \Psi(0,0)\Psi(x,\tau)\rangle$.
To this end, we write:
\begin{eqnarray}
\label{Expansion}
\Psi &=& e^{im\phi_1}\sum\limits_{n=-\infty}^{\infty}
	c_n e^{in(k_0x+\varphi)},
\end{eqnarray}
In the long wave length limit, the dominant contribution to $g(x,\tau)$
will come from the term with $n=0$ in the above series. Therefore,
we only retain this term in what follows; the effect of the omitted
terms will be commented upon in the discussion. Since the action of
system is Gaussian, the evaluation of the Green's function is now
straightforward. The  expressions turn out quite cumbersome, but
one may exploit the limits $v_\varphi/\ll 1$, $x,\tau\to\infty$ to
obtain:
\begin{eqnarray}
\label{Greens_function}
g(x,\tau) \propto \prod\limits_{i=1}^{3}\frac{1}{(x+iv_i\tau)^{\alpha_i}}.
\end{eqnarray}
Here $v_1$, $v_2$ and $v_3$ are the velocities of the three modes:
$v_1\approx v(1-2\beta)$, $v_{2,3}\approx \pm v_\varphi(1+\beta)$,
where $\beta = (\pi K_2^2/mK_1)(v_\varphi/v)$ is small.
To second order in $v_\varphi/v$, the exponents are:
$\alpha_1\approx m$, $\alpha_{2,3}\approx (4\pi K_2^2/K_1)(v_\varphi/v)^2$.
The sum of these three exponents describes local electron tunneling
into the reconstructed edge and has been obtained earlier.~\cite{Yang}
Note also, that the ``velocity conservation law'', $\sum v_i = v$,
derived in~\cite{Heinonen-Eggert} for a similar model, is not satisfied in our
case due to slightly different interaction term $S_{\rm int}$.

Next, one encounters a complication:
the expression for the Green's function Eq.~(\ref{Greens_function})
is, in general, not single-valued. Therefore,
the analytic continuation to the real time, needed to find the spectral
function, is ambiguous and depends on the choice of branch cuts. A
way around this difficulty was found in
Ref.~\onlinecite{Carpentier-Peca-Balents}
for a similar problem, where an elegant trick was used to make the
analytic continuation. However, this trick cannot be used in the present
case because it fails for the values of the exponents $\alpha_i$
that appear in our problem. We shall use a different approach.

Notice that the Green's function in Eq.~(\ref{Greens_function}) is
a product of three factors, each having a singularity corresponding to
one of the three propagating modes after the reconstruction. The Green's
function thus has the form:
\begin{equation}
G = g_1 g_2 g_3,
\end{equation}
where $g_i$ are (fictitious) Green's functions corresponding to the three
modes. But each of these three Green's functions has a form identical
to the electron Green's function of the original (unreconstructed) edge,
the only variation between the three being the signs and absolute values
of the velocities $v_i$ of the edge modes and the edge exponents
$\alpha_i$. This allows us to write the electron spectral function as a
convolution:
\begin{eqnarray}
\label{Spectral_function}
A(\Omega,Q) = \int \prod\limits_{i=1}^3 A_i(\omega_i,q_i)
	[\prod\limits_{i=1}^3\theta(\omega_i)
	+ \prod\limits_{i=1}^3\theta(-\omega_i) ]\nonumber\\
\times \delta(\Omega-\sum\limits_{i=1}^3\omega_i)
	\delta(Q-\sum\limits_{i=1}^3q_i)
	\prod\limits_{i=1}^3 d\omega_i dq_i,
\end{eqnarray}
where $A_i(\omega,q) \propto |q|^{\alpha_i-1}\delta(\omega-v_iq)$ are
the spectral functions corresponding to the Green's functions $g_i$.
A typical plot of the spectral function is shown in Fig.~2.
As a result of edge reconstruction, some
of the spectral weight is shifted away from the original $\delta$-function
singularity at $\omega=vq$ (its new position $v_1q$ is itself slightly
renormalized). An extra pair of singularities appear at
$\omega=v_{2,3} q$; these singularities correspond to the new
neutral modes. An important feature of the spectral function  is a
finite amount of
weight at $\omega<0$. This is possible only due to the fact that,
after the reconstruction, an excitation mode appears that propagates
in the direction opposite to the direction of the original edge
mode.

\insertgraphics{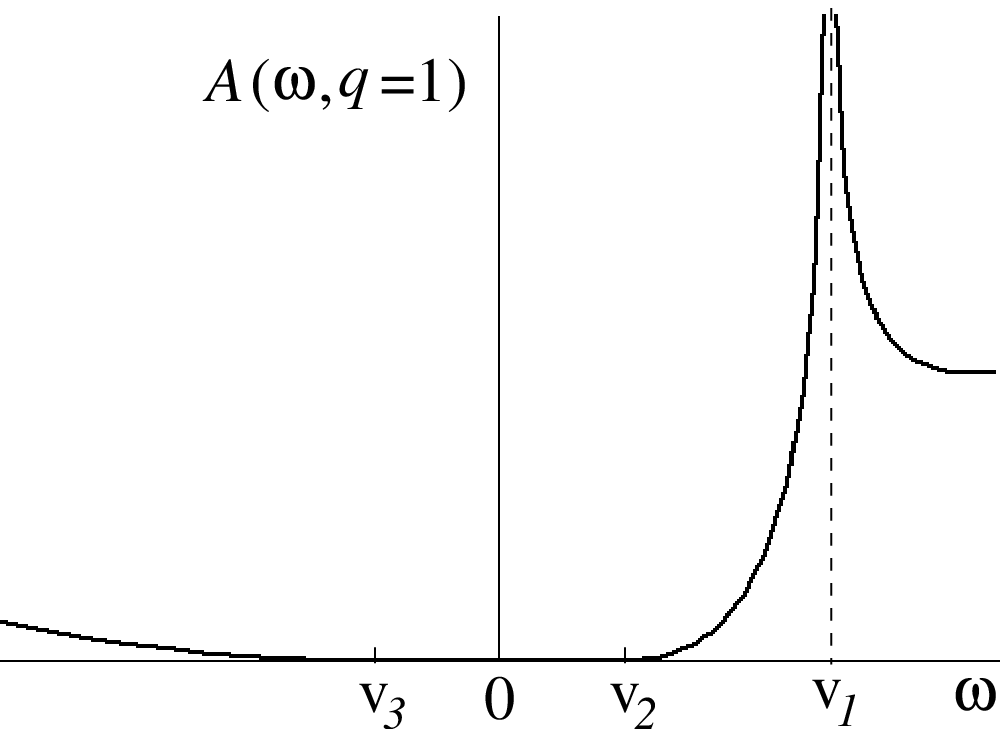}{The electron
spectral function in the reconstructed edge has singularities on
the lines $\omega=v_iq$ that correspond to the edge excitation modes.
The singularity at $\omega=v_1q$ is a divergence, and is the remnant of
the $\delta$-peak in the unreconstructed edge. The spectral function
vanishes for $v_3q<\omega<v_2q$ (e.g. for $q>0$) due to
kinematic constrains. Shown is the case of the filling fraction $\nu=1/3$.}

We now turn to the question of the possibility of experimental
observation of the new features that appear in the electron
spectral function as a result of edge reconstruction. It is clear that
these features are associated with the redistribution of the weight
of the original edge mode. Therefore, one  should look for an
experiment in which
the observed quantity is most closely related to the spectral function
itself and not its integral over $\omega$ or $q$. Such a situation
is realized in the so-called momentum-resolved tunneling
experiments.~\cite{Kang,Grayson} In these experiments, the electron tunneling
into the edge occurs across
a barrier which is extended and homogeneous along the edge of the
FQHE system, and so the electron's momentum along the edge is conserved.
In what follows, we consider the simplest possible case of tunneling
from the (non-reconstructed) edge of the QHE at the filling fraction
$\nu=1$ which is parallel to the reconstructed edge. The spectral function
of the unreconstructed $\nu=1$ edge is just a $\delta$-function. This serves
best to reveal the features of the spectral function of the reconstructed
edge.

\insertgraphics{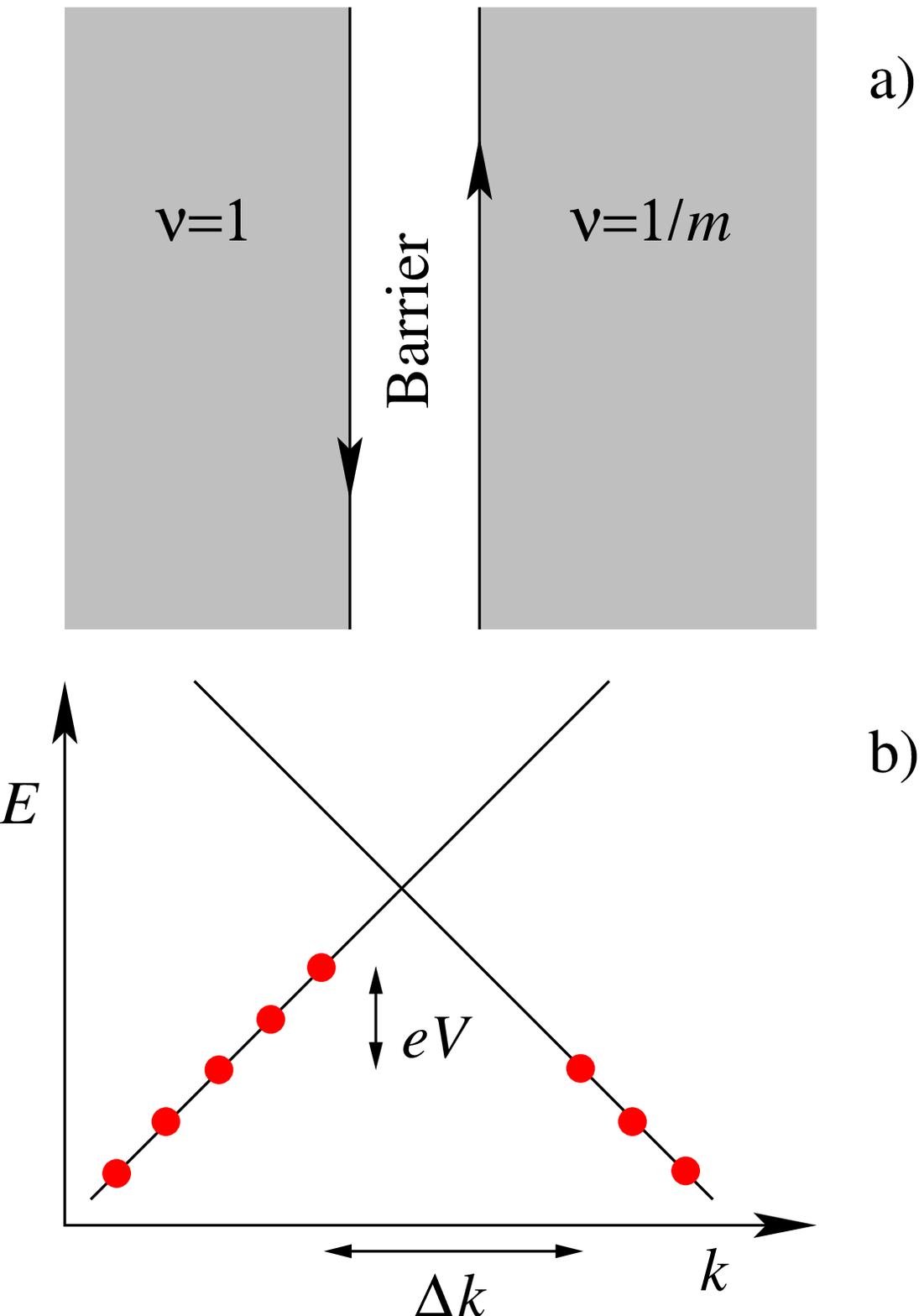}{a) The setup consists of $\nu=1$
and reconstructed $\nu=1/m$ states separated by a barrier which is uniform
along the edges; b) The structure of the energy dispersion in the vicinity of
the barrier (see text for details).}

The setup and the structure of energy levels under these conditions
are shown in Fig.~3. The energy $E$ of the states is plotted as a function of
momentum $k$ along the two parallel edges.
The two straight lines in plot b) correspond to single electron 
dispersion on the two sides of the barrier.
Near the intersection point of these lines, the states
on the left and on the right begin to mix, and tunneling becomes possible.
We shall treat tunneling
using Fermi golden rule. Finally, we neglect the interaction between
the edge modes on the opposite sides of the barrier, while the intra-edge
interactions are taken into account by strong renormalization of the
(charged) mode velocities.

In what follows, we shall consider separately two cases: $m=1$ and $m=3$.
For energetic reasons, the edge of the QH state with higher $m$ is more
susceptible to reconstruction,~\cite{Wan-Yang-Rezayi}
however, the changes in the spectral
function are more noticeable for lower $m$.

There are two parameters that one can control: bias voltage $V$ and 
magnetic field $B$. These two parameters are directly related to the
structure of the spectrum: $eV$ sets the difference between the
Fermi energies on the two sides of the barrier, while the difference
between the two Fermi momenta $\Delta k \propto B-B_0$, where $B_0$ is
the strength of the magnetic field at which, in the absence of bias, the
Fermi energy lies exactly at the dispersion lines crossing point.
Below, we shall rescale $B$ so that $\Delta k = B-B_0$.

Within our approximation, the tunneling current is given by the Fermi
golden rule:
\begin{eqnarray}
\label{Fermi}
I(V,B) &\propto& \int A_1(\omega_1,q_1)A_2(\omega_2,q_2)
	[f(\omega_1)-f(\omega_2)]\nonumber\\
	&&\times\delta(eV+\omega_1-\omega_2)\delta(B-B_0-q_1-q_2)\nonumber\\
	&&\times d\omega_1 d\omega_2 dq_1 dq_2.
\end{eqnarray}
Here, $A_1=\delta(\omega-v_Fq)$ is the spectral function of the $\nu=1$ edge
with the Fermi velocity $v_F$, and $A_2$ is the spectral function in
Eq.~(\ref{Spectral_function}); $f(\omega)$ is the Fermi distribution
step function.

\insertwidegraphics{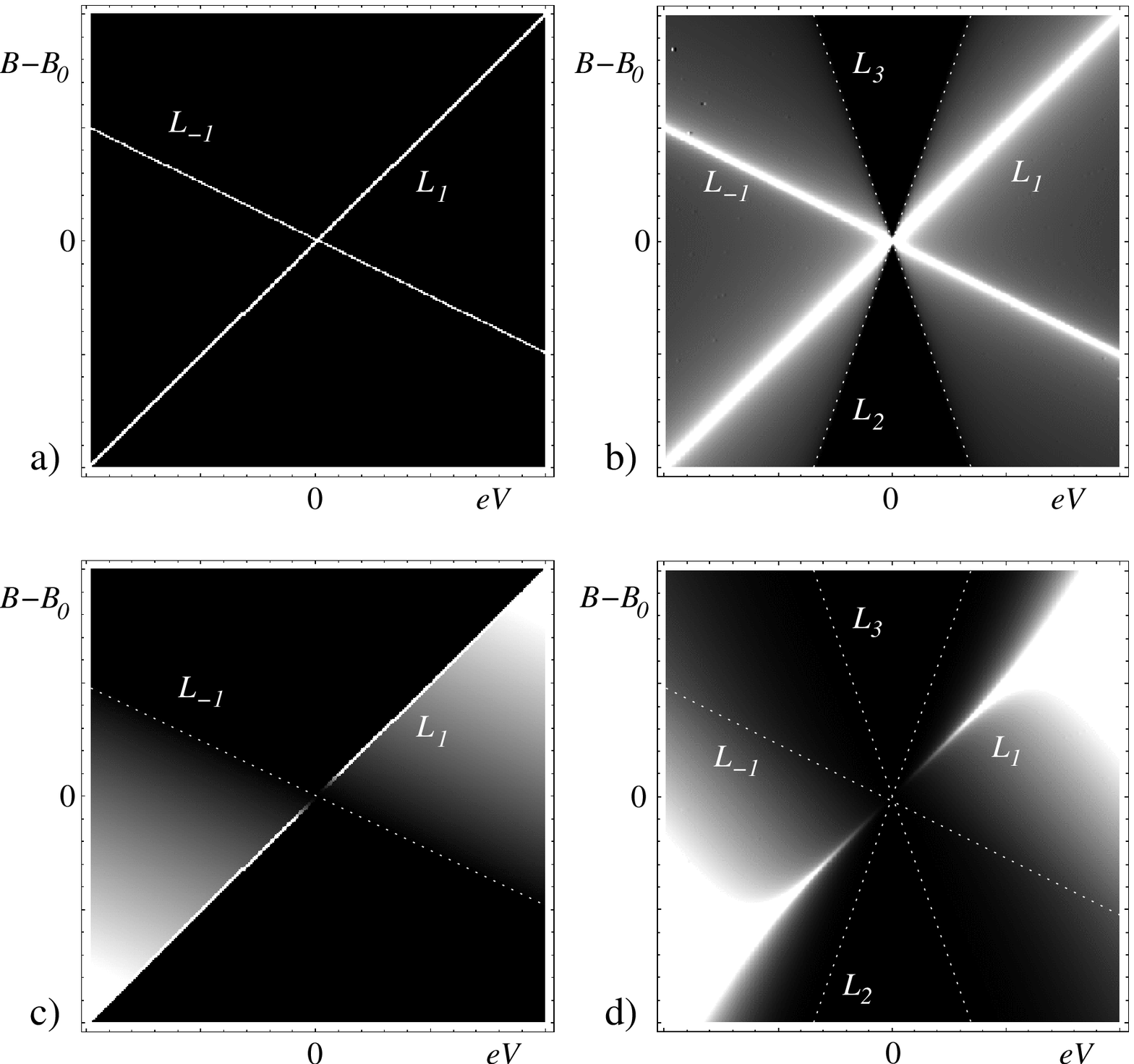}{Differential conductance $dI/dV$
as a function of bias voltage $V$ and magnetic field $B$
for momentum-resolved tunneling from unreconstructed $\nu=1$ edge to:
a) unreconstructed $\nu=1$ edge; b) reconstructed $\nu=1$ edge;
c) unreconstructed $\nu=1/3$ edge; d) reconstructed $\nu=1/3$ edge.
Plotted is the derivative of the result in Eq.~(\ref{Fermi}). Parameters
are: $v_1=1$, $v_{2,3}=\pm 1/3$, $v_F=2$, $\alpha_1=m=\nu^{-1}$,
$\alpha_{2,3}=0.25$. The dashed lines are guides to the eye and are defined
as $L_{1,2,3}=\{eV=v_{1,2,3}(B-B_0)\}$, $L_{-1}=\{eV=-v_F(B-B_0)\}$.
See text for details.}

The result of the numerical evaluation
of the differential conductance $dI/dV$ as a function of $V$ and $B-B_0$
is shown in Fig.~4. In general, the plot of
$dI(V,B)/dV$ is very similar, but not identical to $A_2(\omega,q)$.
In particular, both have 3 lines of singularities (marked by letters
$L_{1,2,3}$ on the figure), one of them being a divergence. These lines
correspond to the three edge excitation modes $\omega=v_iq$. Moreover,
in the bow-tie region between $L_2$ and $L_3$ the differential conductance
(as well as the current itself) is exactly zero. The reason for this is
purely kinematic: In the region below the dispersion line
of the slowest excitation in the system, one cannot satisfy the conservation
of energy and momentum in tunneling. This is best visible if one plots
$dI/dV$ in a sweep of $V$ at fixed $B$. The resulting plot is very similar
to that of the spectral function in Fig.~2. In the situation, where
reconstruction produces multiple edges, this kinematic constraint
dictates that $dI/dV=0$ in the region set by the two {\em slowest}
velocities in the system.

We would like to point out,
that while the singularities that correspond to the new neutral modes
are rather weak and thus are difficult to observe, the general structure
of the spectral weight transfer provides a clear indication
of the edge reconstruction. In particular, in the bow-tie regions between
$L_1$ and $L_2$, and between $L_3$ and $L_{-1}$, $dI/dV$ is zero before
reconstruction and finite after. We note, that the rise of $dI/dV$
in the region between $L_3$ and $L_{-1}$ is due to the appearance of
a (neutral) edge mode that propagates in the direction opposite
to the direction of the original edge mode.

Although we have concentrated on the case of edge
reconstruction described by a single point $k_0$, the preceding discussion
remains generally valid for multiple edge reconstruction as well.
In that case, new lines of singularities appear, while $L_2$ and $L_3$
correspond to the two slowest excitation modes.

A careful analysis~\cite{Progress} allows one to extract the precise form
of the diverging singularities on $L_1$ and $L_{-1}$. The singularity on
$L_1$ has the form: $[eV-v_1(B-B_0)]^{\alpha_2+\alpha_3-1}$.
The singularity on $L_{-1}$ has the form:
$[eV+v_F(B-B_0)]^{\alpha_1+\alpha_2+\alpha_3-2}$. In particular, the latter
result explains why, given the Coulomb-interaction induced smallness of
the exponents $\alpha_{2,3}$, the singularity on $L_{-1}$ in the $\nu=1/3$
case with $\alpha_1\approx 1/\nu = 3$ (Fig.~4.d) is non-divergent. In
principle, one should be able to extract independently the values of
$\alpha_1$ and $\alpha_2+\alpha_3$ by fitting the experimental data to the
above expressions, thus offering a consistency check,
$\alpha_1\approx 1/\nu$, and a measure of the ``strength'' of
reconstruction, $\alpha_2+\alpha_3$. Finally, the combination
$\alpha_1+\alpha_2+\alpha_3$ is the exponent that describes the local
tunneling into the reconstructed quantum Hall edge.~\cite{Yang}
Momentum-resolved tunneling experiments thus offer an independent way
of measuring this exponent.


Finally, we would like to comment on the role of the omitted $n\neq 0$
terms in Eq.~(\ref{Expansion}). Apart from the factors $\exp ink_0x$, which
trivially shift the momentum argument of the electron spectral function
by $nk_0$, all terms with $n \neq 0$ are qualitatively similar to the
$n=0$ term. Each of them produces a contribution to the Green's function
which is of the from Eq.~(\ref{Greens_function}), albeit with
larger exponents $\alpha_i$. For that reason, all these terms make
progressively less visible (but not necessarily unobservable)
contributions to the spectral function.

Summarizing, we find that that edge reconstruction qualitatively changes
the electron spectral function at the edge. The weight of the original
sharp peak undergoes broad redistribution and the new edge modes that appear
after the reconstruction show up as lines of singularities in the
electron spectral function. We also suggest a momentum-resolved
tunneling experiment which is best suited for probing the predicted
features. We find that some of these features can be used to infer
edge reconstruction.


We thank Matt Grayson, Woowon Kang and Leon Balents for helpful discussions.
This work was supported by NSF grant No. DMR-0225698.


\end{document}